\documentclass[letterpaper, 10 pt, conference]{ieeeconf}  

\IEEEoverridecommandlockouts                 

\makeatletter

\let\proof\@undefined
\let\endproof\@undefined
\makeatother

\usepackage{amsmath,amssymb,mathabx}
\usepackage{amsfonts,mathtools,amsthm}
\usepackage{mathrsfs,bm,dsfont}
\makeatletter
\newcommand{\stepsubequations}{%
  \ifmeasuring@
  \else
    \stepcounter{parentequation}\setcounter{equation}{0}%
    \xdef\theparentequation{\arabic{parentequation}}%
  \fi
}
\makeatother
\newcommand{\colvec}[2][.9]{%
  \scalebox{#1}{%
    \renewcommand{\arraystretch}{1}%
    $\begin{bmatrix}#2\end{bmatrix}$%
  }
}

\newtheorem*{remark}{Remark}

\usepackage[dvipsnames]{xcolor}
\usepackage{subcaption}
\usepackage{multirow}
\usepackage{booktabs}
\usepackage{pgf}
\usepackage[normalem]{ulem}

\usepackage{tikz}
\definecolor{myred}{rgb}{0.65,0.16,0.16}

\DeclareCaptionLabelSeparator{periodspace}{.\quad}
\usepackage{titlesec}
\titleformat{\subsubsection}[runin]{\normalfont\normalsize\it}{\thesubsubsection}{}{}[~-]
\titlespacing*{\subsubsection}{2pt}{7pt}{0.5em}
\titlespacing*{\section}{2pt}{4pt}{4pt}

\usepackage{hyperref}
\hypersetup{%
   colorlinks = true,%
   citecolor = TealBlue,%
   urlcolor  = TealBlue,%
   linkcolor = MidnightBlue%
}

\usepackage[
placement=top,
angle=0,
color=red,
scale=1.1,
hshift=0,
vshift=-15
]{background}
\backgroundsetup{contents=\bf{----------------------------------------------------------------- PREPRINT -----------------------------------------------------------------}}

\title{\LARGE \bf
Star-shaped Tilted Hexarotor Maneuverability:\\ Analysis of the Role of the Tilt Cant Angles}

\author{Marco Perin$^{1,3}$, Massimiliano Bertoni$^{2}$, Nicolas Viezzer$^{1}$, Giulia Michieletto$^{1,2}$ and Angelo Cenedese$^{1}$
 \thanks{*This work is partially supported by MUR through PRIN Grant DOCEAT 2020RTWES4 and by the European Union – Next Generation Eu - under the National Recovery and Resilience Plan (NRRP), Mission 4 Component 1 Investment 4.1  - Call for tender No. 2152 of Italian Ministry of University and Research; Concession Decree No. 2152 adopted by the Italian Ministry of University and Research, Project code D93C22000850005, within the Italian National Program PhD Programme in Autonomous Systems (DAuSy).}
 \thanks{$^{1}$Dept. of Information Eng., University of Padova, Italy.
        }%
 \thanks{$^{2}$Dept. of Management and Eng., University of Padova, Italy.
        }%
 \thanks{$^{3}$Dept. of Electrical and Information Eng., Polytechnic of Bari, Italy 
 }%
\thanks{{{Contacts: M.Perin - 
 {\tt\small marco.perin.6@studenti.unipd.it}}
 }
         }%
}

\begin{document}

\maketitle
\thispagestyle{empty}
\pagestyle{empty}

\begin{abstract} 
Star-shaped Tilted Hexarotors are rapidly emerging 
for applications highly demanding in terms of robustness and maneuverability. To ensure improvement in such features, a careful selection of the tilt angles is mandatory. 
In this work, we present a rigorous analysis of how the force subspace varies with the tilt cant angles, namely the tilt angles along the vehicle arms, taking into account gravity compensation and torque decoupling to abide by the hovering condition. 
Novel metrics are introduced to assess the performance of existing tilted platforms, as well as to provide some guidelines for the selection of the tilt cant angle in the  design phase.

\end{abstract}

\section{Introduction}

In the last decade, Unmanned Aerial Vehicles (UAVs) have drawn growing interest within the robotics community due to the unique challenges they present in designing control and estimation solutions, alongside their outstanding versatility across diverse applications, ranging from conventional monitoring operations to modern physical interaction tasks~\cite{mohsan2023unmanned}.

The demand for efficient solutions in cutting-edge aerial robotics has led to the development of new UAV configurations. These offer enhanced actuation capabilities compared to traditional quadrotors, which are limited by under-actuation and coupled dynamics resulting from their standard coplanar and collinear arrangement of actuators.
Interest has grown in the design of fully-actuated UAVs, equipped with six actuators arranged to ensure full controllability. 
Moreover, literature has recently focused on multi-rotor UAVs featuring tilted or tilting propellers. Tilted propellers have a fixed orientation relative to the vehicle's local frame, tilting propellers can adjust their orientation during flight, allowing for improved flight capabilities at the cost of a more complex design and control architecture~\cite{rashad2020fully}. 

Among all the state-of-the-art UAVs, the star-shaped tilted hexarotors (hereafter, we refer to them as STHs) represent the class of aerial platforms with the simplest design yet ensuring full actuation and complete force-moment decoupling.
STHs are characterized by six actuators placed at the vertices of a hexagon centered on the vehicle center of mass (CoM) and consisting of statically suitably tilted propellers.
These structural features entail the possibility to 
allocate the control force and moment in three-dimensional domains, making this type of UAV beneficial in various application contexts, as, for instance, for operations in harsh and highly disturbed environments and contact-aware interacting tasks (see e.g.~\cite{lee2018design}).

\subsubsection*{Related Works}
Due to their emerging high potential, STHs have become the subject of numerous studies focusing on optimizing platform design, implementing efficient estimation and control solutions, and analyzing their dynamic characteristics (see e.g.,~\cite{rajappa2015modeling,yao2018modeling} and the references therein). As concerns this last aspect, twofold features are generally investigated for these UAVs: their capability to realize the static hovering condition even in the presence of an actuator failure~\cite{giribet2016analysis,mochida2021geometric} and the peculiarities of the control force and control moment domains, as well as their interplay, depending on the tilt angles~\cite{mehmood2016maneuverability,michieletto2019force,kotarski2021performance}. Specifically, in~\cite{mehmood2016maneuverability} the STH maneuverability is analyzed based on the maximum acceleration achievable in a particular direction of the 3D space; in~\cite{michieletto2019force} the force-moment decoupling is evaluated by accounting for the spectral analysis of the control input matrices;
in~\cite{kotarski2021performance} the UAV performance is discussed through the definition of a control allocation scheme that describes the STH configuration in terms of tilt angles.

\subsubsection*{Contributions}
From the existing literature, it emerges that tilting the rotors around the axes coinciding with the vehicle arms (non-zero tilt cant angles) has a more significant effect in terms of improving maneuverability compared to tilting around the axes perpendicular to the vehicle arms (non-zero dihedral angles), which instead benefits the possibility of performing static hovering in case of a propeller stop. For these reasons, in this work, the attention is focused on the role of cant angles, rather than dihedral angles. Specifically, we focus on platforms whose rotors are alternately equally tilted. For this class of UAVs, we outline a rigorous method to investigate the maneuverability properties as a function of the cant angles selection. This is based on the polytope of the feasible control forces ensuring torque decoupling up to gravity compensation. In more detail, we define some ad-hoc geometrically inspired metrics that can be used both to analyze the capabilities of existing platforms and to guide the design of new STH configurations.

\subsubsection*{Paper structure}
The rest of the paper is organized as follows. Sec.~\ref{sec:preliminaries} presents some preliminaries on the used models. Sec.~\ref{sec:main_force} is devoted to the discussion of maneuverability properties while Sec.~\ref{sec:discussion} depicts an example of design choice. Finally, Sec.~\ref{sec:conclusions} summarizes the main conclusions and draws future research steps.


\section{Preliminaries}
\label{sec:preliminaries}

\begin{figure}[t!]
    \centering
    \includegraphics[trim={0 0 0 3cm},clip,width=0.9\columnwidth]{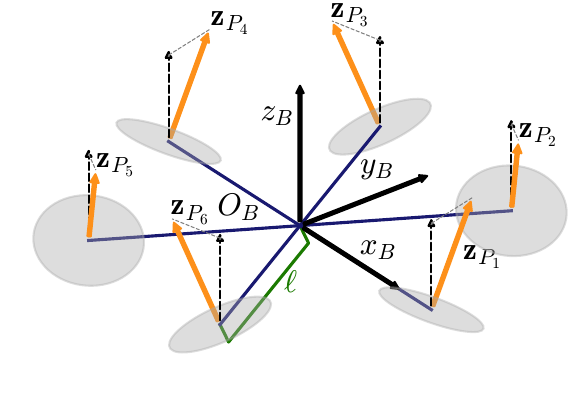}
    \caption{STH model - yellow arrow represent the tilted rotors spinning axes.}
    \label{fig:hexa_model}
\end{figure}

This section focuses on the STH actuation model, applying the force analysis from~\cite{michieletto2019force} to the study case in Fig.~\ref{fig:hexa_model}.

\subsection{STH Actuation Model}

We refer to STH as a multi-rotor platform actuated by six propellers, each of which is placed in correspondence with a vertex of a regular hexagon centered on the UAV CoM and is possibly tilted around the axis corresponding to the vehicle arm. In particular, in this work, we restrict the attention to the STHs wherein the tilt angles cannot vary during flight, and adjacent propellers are alternatively tilted at the same angle. Hence, each STH is identified by a specific selection of the \textit{tilt cant angle} $\alpha \in \Gamma_\alpha:=\left[ 0, 90^\circ \right)$. 
Note that the star-shaped collinear hexarotors characterized by the selection $\alpha=0$  are considered a particular realization of STHs.

We introduce the global inertial reference frame $\mathscr{F}_W =\{O_W,(\mathbf{x}_W,\mathbf{y}_W,\mathbf{z}_W)\}$ (\textit{world frame}) whose axes directions are identified by the unit vectors $\mathbf{e}_1$, $\mathbf{e}_2$ and $\mathbf{e}_3$ of the canonical basis of $\mathbb{R}^3$ for sake of simplicity, and the local reference frame $\mathscr{F}_B=\{O_B,(\mathbf{x}_B,\mathbf{y}_B,\mathbf{z}_B)\}$ (\textit{body frame}), in-built with the vehicle and centered in its CoM. 
The position and orientation of an STH in the 3D space are thus described by the vector $\mathbf{p}\in\mathbb{R}^3$ denoting the position of $O_B$ in $\mathscr{F}_W$, and by the rotation matrix $\mathbf{R} \in SO(3)$  representing the orientation of $\mathscr{F}_B$ with respect to  $\mathscr{F}_W$. 
Then, under the star-shaped hypothesis, the position of any $i$-th propeller, $i \in \{1 \ldots 6\}$, in $\mathscr{F}_B$, is described by the vector $\mathbf{p}_i \in \mathbb{R}^3$,
\begin{align}
\label{eq:positions_gamma}
\mathbf{p}_i &= \ell \,
\mathbf{R}_z \left((i-1) 60^\circ \right) \mathbf{e}_1, 
\end{align}
where $\ell \in \mathbb{R}_{> 0}$ is the distance between $O_B$ and the propeller CoM assumed on the $(\mathbf{x}_B,\mathbf{y}_B)$ plane.
In addition, under the tilt hypothesis, the orientation of the spinning axis of any $i$-th propeller is identified by the unit vector $\mathbf{z}_{P_i}(\alpha) \in \mathbb{R}^3$,
\begin{equation}
\label{eq:orientations_alphabeta}
\mathbf{z}_{P_i}(\alpha) =\mathbf{R}_z  \left((i-1) 60^\circ\right)\mathbf{R}_x((-1)^{i}\alpha)\mathbf{e}_3.
\end{equation}
 In~\eqref{eq:positions_gamma}-\eqref{eq:orientations_alphabeta}, the matrices $\mathbf{R}_x(\cdot)$ and $\mathbf{R}_z(\cdot) \in SO(3)$ represent elemental rotations around the $x$ and $z$ axes, respectively.

By rotating around its spinning axis, each $i$-th propeller generates in its CoM a \textit{thrust force} $\mathbf{f}_i(\alpha) \in \mathbb{R}^3$ and a \textit{drag moment} $\boldsymbol{\tau}_i^d(\alpha)\in \mathbb{R}^3$ having constant direction in $\mathscr{F}_B$ depending on the tilt angle.
Both the thrust force and the drag moment, together with the emerging \textit{thrust moment} $\boldsymbol{\tau}_i^t(\alpha) = \mathbf{p}_i \times \mathbf{f}_i(\alpha) \in \mathbb{R}^3$, 
depend on the propeller
spinning rate $\omega_i \in \mathbb{R}_{\geq 0}$
according to the popular models
\begin{align}
\mathbf{f}_i(\alpha)  &=    c_{f_i} \omega_i^2\mathbf{z}_{P_i}(\alpha) ,
\label{eq:thrust_force}\\
 {\boldsymbol\tau}_i^d(\alpha)  &= \kappa_i c_{\tau_i} \omega_i^2\mathbf{z}_{P_i}(\alpha) ,
\label{eq:drag_moment}\\
\boldsymbol{\tau}_i^t(\alpha)  &=  c_{f_i} \omega_i^2  (\mathbf{p}_i \times \mathbf{z}_{P_i}(\alpha) )  \label{eq:thrust_moment}
\end{align}
where  $c_{f_i},c_{\tau_i} \in \mathbb{R}_{\geq 0}$ are constant parameters depending on the rotor geometric features and $\kappa_i \in \{-1,1\}$ allows for distinguishing whether   the $i$-th propeller spins counterclockwise (CCW, $\kappa_i=1$) or clockwise (CW, $\kappa_i=-1$). 

\begin{figure*}[t!]
\setcounter{equation}{11}
\begin{subequations} 
\label{eq:FM}
     \begin{align}
    &\mathbf{F}_\alpha=  c_{f}  \colvec[0.9]{
    0 & \frac{\sqrt{3}}{2}s\alpha & -\frac{\sqrt{3}}{2}s\alpha & 0 & \frac{\sqrt{3}}{2}s\alpha & -\frac{\sqrt{3}}{2}s\alpha\\
    s\alpha & -\frac{1}{2} s\alpha &  - \frac{1}{2} s\alpha & s\alpha & -\frac{1}{2} s\alpha & - \frac{1}{2} s\alpha\\
    c\alpha  & c\alpha  & c\alpha  & c\alpha & c\alpha & c\alpha } \\
    &\mathbf{M}_\alpha= c_{\tau} \colvec[0.9]{
   0 & \frac{\sqrt{3}}{2} r \, c\alpha -\frac{\sqrt{3}}{2}s\alpha & \frac{\sqrt{3}}{2}r \, c\alpha -\frac{\sqrt{3}}{2} s\alpha & 0 & -\frac{\sqrt{3}}{2} r \, c\alpha  + \frac{\sqrt{3}}{2} s\alpha & -\frac{\sqrt{3}}{2} r \, c\alpha +\frac{\sqrt{3}}{2} s\alpha\\
   -r \, c\alpha +s\alpha & -\frac{1}{2} r \, c\alpha  + \frac{1}{2} s\alpha & \frac{1}{2}r \, c\alpha  -\frac{1}{2}s\alpha & r \, c\alpha -s\alpha &\frac{1}{2} r \, c\alpha -\frac{1}{2} s\alpha & -\frac{1}{2}r \, c\alpha +\frac{1}{2}s\alpha\\
   r \, s\alpha +c\alpha  & -r \, s\alpha -c\alpha  & r \, s\alpha +c\alpha  & - r \, s\alpha - c\alpha  & r \, s\alpha +c\alpha  & - r \,s\alpha -c\alpha  
   } \end{align}
       \end{subequations}
\hrule
\end{figure*}

Considering all propeller actions
regulated through the assignable \textit{control input} $u_i= \omega_i^2 \in \mathbb{R}_{\geq 0}$,   
the \textit{total control force} $\mathbf{f}_{c}(\alpha) \in \mathbb{R}^3$ and the \textit{total control moment} ${\boldsymbol\tau}_{c}(\alpha)$ applied in the platform CoM and expressed in $\mathscr{F}_B$ result to be 
\setcounter{equation}{8}
\begin{align}
\mathbf{f}_{c}(\alpha) &=  {\textstyle\sum_{i=1}^6} \mathbf{f}_i(\alpha)  \label{eq:thrust_control} \\
&= 
 {\textstyle\sum_{i=1}^6} c_{f_i}   \mathbf{z}_{P_i}(\alpha) u_i, \nonumber
\\
{\boldsymbol\tau}_{c}(\alpha) &= {\textstyle\sum_{i=1}^6} (\boldsymbol{\tau}^t_i(\alpha)+{\boldsymbol\tau}_i^d(\alpha)) \label{eq:moment_control}  \\ 
&
= {\textstyle\sum_{i=1}^6} 
\left(c_{f_i}\mathbf{p}_i \times  \mathbf{z}_{P_i}(\alpha) 
+ \kappa c_{\tau_i}\mathbf{z}_{P_i}(\alpha)
\right) u_i. \nonumber
\end{align}
Introducing the \textit{control input vector} $\mathbf{u} = \colvec{ u_1 \; \cdots \; u_6 }^\top \in\mathbb{R}^6$, expressions~\eqref{eq:thrust_control} and~\eqref{eq:moment_control} can be
shortened to become
\begin{align}
\mathbf{f}_{c}(\alpha)=\mathbf{F}_\alpha\mathbf{u} \quad \text{and} \quad \boldsymbol{\tau}_{c}(\alpha)= \mathbf{M}_\alpha\mathbf{u}, 
\end{align}
where the \textit{control force input matrix} $\mathbf{F}_\alpha\in \mathbb{R}^{3 \times 6}$ and the \textit{control moment input matrix}
$\mathbf{M}_\alpha\in \mathbb{R}^{3 \times 6}$ depend on the tilt angle and on the geometric and aerodynamic parameters of the propellers.  In the rest of the paper, we account for platforms actuated by a set of rotors with equal actuation and  aerodynamic characteristics, as well as a balanced choice of CW/CCW spinning directions.
Specifically,  for $i \in \{1 \ldots  6\}$, we assume that $u_i \in \bar{\mathcal{U}} = [0,\bar{u}]$ with $\bar{u} \in \mathbb{R}_{\geq 0}$, $c_{f_i} = c_f$, $c_{\tau_i}=c_\tau$ with $c_f>c_\tau$, and $\kappa_i=(-1)^{i}$. 
The matrices $\mathbf{F}_\alpha$ and 
$\mathbf{M}_\alpha$ thus result as in~\eqref{eq:FM} where $r =  (c_f/c_\tau) \ell \in \mathbb{R}_{\geq 0}$ and $c$ and $s$ stand for cosine and sine function, respectively. 

\begin{remark}
    In the rest of the paper, the numerical results refer to a case study of the STH platform described in~\cite{bertoni2022indoor} and characterized by the parameters reported in Tab.~\ref{tab:STH_parameters}. 
   \end{remark}

\begin{table}[h!]
    \centering
    \begin{tabular}{ccccc}
    \toprule
        $m$[kg] &  $\ell$ [m] & $c_f$[N/Hz$^2$] & $c_{\tau}$[Nm/Hz$^2$]  & $\bar{u}$ [Hz$^2$]  \\\midrule
         $3.500$ & $0.385$ & $1.500e-03$ & $4.590e-05$ &  $108^2$ \\ \bottomrule
    \end{tabular}
    \caption{Parameters of the STH platform case study}
    \label{tab:STH_parameters}
\end{table}

For any choice of tilt angle $\alpha \in \Gamma_\alpha$, it follows that $1 \leq \text{rk}\mathbf{F}_\alpha\leq \text{rk}\mathbf{M}_\alpha\leq 3$ as explained in~\cite{michieletto2019force}. In detail, the control moment input matrix is always full rank when $\alpha \in \Gamma_\alpha$,
while the control force input matrix is rank deficient when $\alpha=0$, i.e., for collinear hexarotors (note also that $\text{rk}\mathbf{F}_\alpha=2$ when $\alpha= \pm 90^\circ$ corresponding to the case wherein the thrust force generated by any propeller lies on the  $(\mathbf{x}_B,\mathbf{y}_B)$ plane).
Finally, a STH is fully actuated if $\text{rk}\mathbf{C}_\alpha= 6$ being $\mathbf{C}_\alpha= \colvec{\mathbf{F}_\alpha^\top \; \mathbf{M}_\alpha^\top}^\top \in \mathbb{R}^{6\times 6}$. This condition is satisfied if $\alpha\in \Gamma_\alpha \backslash \{ 0\}$.


Adopting the Euler-Newton approach, 
the dynamics of the platform can be described through the following equations 
\setcounter{equation}{12}
 \begin{align}
  m\ddot{\mathbf{p}} &= - mg \mathbf{e}_3 +  \mathbf{R}\mathbf{f}_c(\alpha)  
  =- mg \mathbf{e}_3 +  \mathbf{R}\mathbf{F}_\alpha \mathbf{u},
 \label{eq:pos_dyn}\\
  \mathbf{J} \dot{\boldsymbol{\omega}} &= -  \boldsymbol{\omega} \times \mathbf{J} \boldsymbol{\omega} 
 + {\boldsymbol\tau}_{c}(\alpha) 
 =-  \boldsymbol{\omega} \times \mathbf{J} \boldsymbol{\omega} +\mathbf{M}_\alpha \mathbf{u}, \label{eq:orient_dyn}
 \end{align}
 where $g \in \mathbb{R}_{> 0}$ and $m \in \mathbb{R}_{> 0}$ are the gravity constant and the total platform mass, while $\mathbf{J} \in \mathbb{R}^{3 \times 3}$ is the positive definite inertia matrix of the vehicle computed in $\mathscr{F}_B$.

\subsection{Control Force Decomposition}
\label{sec:force_decomposition}

From~\cite{michieletto2019force}, introducing the full-rank matrices $\mathbf{A}_\alpha \in \mathbb{R}^{6\times 3}$ and $\mathbf{B}_\alpha \in \mathbb{R}^{6\times 3}$ so that $\text{Im} (\mathbf{A}_\alpha) = \text{Im}( \mathbf{M}_\alpha^\top)$ and $\text{Im} (\mathbf{B}_\alpha) = \ker (\mathbf{M}_\alpha)$, it is possible to express any control input vector $\mathbf{u} \in \mathcal{U} = \bar{{\mathcal U}}^6$ as the sum of two components, namely
\begin{equation}
\label{eq:input_decomposition}
\mathbf{u} = \mathbf{u}_A + \mathbf{u}_B = \colvec{
\mathbf{A}_\alpha & \mathbf{B}_\alpha} \colvec{\tilde{\mathbf{u}}_A \\ \tilde{\mathbf{u}}_B},  \quad \tilde{\mathbf{u}}_A, \tilde{\mathbf{u}}_B \in \mathbb{R}^3,
\end{equation}
where $\mathbf{u}_A \in \mathcal{U}_A = \mathcal{U} \!\cap\! \text{Im} (\mathbf{A}_\alpha)$ and $\mathbf{u}_B \in \mathcal{U}_B= \mathcal{U}\! \cap \!\text{Im} (\mathbf{B}_\alpha)$.

The decomposition~\eqref{eq:input_decomposition} in turns implies the decomposition of any control force vector.
Specifically, denoting with 
$\mathcal{F}(\alpha) = \left \{\mathbf{f}_c \in \mathbb{R}^3 \; \vert \; \mathbf{f}_c = \mathbf{F}_\alpha \mathbf{u}, \mathbf{u} \in \mathcal{U}\right\} \subseteq \text{Im}(\mathbf{F}_\alpha)$ the \textit{control force space}, any $\mathbf{f}_c\in \mathcal{F}(\alpha)$ can be expressed as the sum of
\begin{align}
\label{eq:force_decomposition}
\begin{aligned}
&\mathbf{f}_c^A(\alpha)=\mathbf{F}_\alpha \mathbf{u}_A \in \mathcal{F}_A(\alpha)\subseteq \text{Im}(\mathbf{F}_\alpha\mathbf{A}_\alpha), \\
&\mathbf{f}_c^B(\alpha)=\mathbf{F}_\alpha \mathbf{u}_B \in \mathcal{F}_B(\alpha) \subseteq \text{Im}(\mathbf{F}_\alpha\mathbf{B}_\alpha)  \quad \text{where}
\end{aligned}  \\
\begin{aligned}
&\mathcal{F}_A(\alpha)=\left \{\mathbf{f}_c \in \mathbb{R}^3 \; \vert \; \mathbf{f}_c = \mathbf{F}_\alpha \mathbf{u}_A , \mathbf{u}_A \in \mathcal{U}_A\right\},\\
&\mathcal{F}_B(\alpha)=\left \{\mathbf{f}_c \in \mathbb{R}^3 \; \vert \; \mathbf{f}_c = \mathbf{F}_\alpha \mathbf{u}_B, \mathbf{u}_B \in \mathcal{U}_B\right\}. 
\end{aligned}
\end{align}
From~\eqref{eq:input_decomposition} it follows also that $\boldsymbol{\tau}_c(\alpha)=\mathbf{M}_\alpha \mathbf{u}_A$, hence $\mathbf{f}_c^A(\alpha)$ represents the `spurious' force arising from the requirement of achieving a certain control moment, while $\mathbf{f}_c^B(\alpha)$ corresponds to the force that can be independently assigned. 

According to~\cite{michieletto2019force}, a UAV is fully decoupled when $\mathbf{f}_c(\alpha)$ can be assigned in $\mathcal{F}(\alpha)$ independently on  $\boldsymbol{\tau}_c(\alpha)$, i.e., when the \textit{zero-moment control force space} $\mathcal{F}_B(\alpha)$ coincides with $\mathcal{F}(\alpha)$. 
For the STHs, $\mathcal{F}_A(\alpha) = \{ \mathbf{0}_3 \}$ and  $\mathcal{F}_B(\alpha)= \mathcal{F}(\alpha)$
for any $\alpha \in \Gamma_\alpha$: any STH is fully decoupled regardless of the tilt angle selection. However, the choice of $\alpha$ affects   $\mathcal{F}_B(\alpha) = \mathcal{F}(\alpha)$, thus influences the platform maneuverability.

\section{Maneuverability Analysis}
\label{sec:main_force}

In this section, we devise a geometric strategy for evaluating STH maneuverability by assessing its capacity to generate arbitrary control forces independently of control moments while ensuring gravity compensation.
Our objectives are twofold: to provide an analysis approach for evaluating an STH's suitability for specific tasks by leveraging its actuation constraints, and to establish design guidelines for new STHs that meet specific maneuverability requirements.

To ease the notation, hereafter we drop the dependence on the tilt angle $\alpha$ when not necessary. Moreover, we summarize all newly introduced symbols within Tab.~\ref{tab:nomenclature}.

\begin{table}[h!]
\centering
\resizebox{0.98\columnwidth}{!}{
\begin{tabular}{ll}
\toprule
\textbf{Symbol} & \textbf{Meaning} \\
\midrule
$\alpha \in \Gamma_\alpha=\left[ 0, 90^\circ\right)$ & tilt cant angle \\[0.1cm]
$\mathbf{F}_\alpha\in \mathbb{R}^{3 \times 6}$ & control force input matrix \\
$\mathbf{M}_\alpha\in \mathbb{R}^{3 \times 6}$ & control moment input matrix \\
$\mathbf{A}_\alpha \in \mathbb{R}^{6\times 3}$ & matrix s.t. $\text{Im} (\mathbf{A}_\alpha) = \text{Im}( \mathbf{M}_\alpha^\top)$\\
$\mathbf{B}_\alpha \in \mathbb{R}^{6\times 3}$ &  matrix s.t. $\text{Im} (\mathbf{B}_\alpha) = \ker (\mathbf{M}_\alpha)$\\
$\mathbf{H}_\alpha \in \mathbb{R}^{3 \times 3}$ & matrix s.t.  $\mathbf{H}_\alpha = \mathbf{F}_\alpha\mathbf{B}_\alpha$ \\[0.1cm]
$\mathcal{{U}} = \bar{\mathcal{U}}^6, \bar{\mathcal{U}}=[0, \bar{u}]$ & space of control input vector, $\bar{u} \in \mathbb{R}_{\geq 0}$ \\[0.1cm]
$\mathcal{F}(\alpha) \in \mathbb{R}^{3}$ & control force (ctrl frc) space  \\
$\mathcal{F}_B(\alpha) \in \mathbb{R}^{3}$ & zero-moment (zm) ctrl frc space \\
$\mathcal{F}_B^h(\alpha) \in \mathbb{R}^{3}$ & zm ctrl frc space with gravity  compensation \\
$\text{conv}(\cdot)$ & convex hull operator \\[0.1cm]
$V_{\mathcal{F}_B} \in \mathbb{R}_{\geq 0}$ & volume of $\text{conv}(\mathcal{F}_B)$ \\
${A}_{\mathcal{F}_B^h} \in \mathbb{R}_{\geq 0}$ & area of $\text{conv}(\mathcal{F}_B^h)$ \\
$V_{\mathcal{F}_B^h} \in \mathbb{R}_{\geq 0}$ & extra-hovering frc ctrl volume \\
$r_{o} \in \mathbb{R}_{\geq 0}$ & outer circle radius of $\text{conv}(\mathcal{F}_B^h)$  \\
$r_{i} \in \mathbb{R}_{\geq 0}$ & inner circle radius of $\text{conv}(\mathcal{F}_B^h)$ \\
\bottomrule
\end{tabular}
}
\caption{Nomenclature used in the paper.}
\label{tab:nomenclature}
\end{table}

\begin{figure*}[t!]
    \centering
    \begin{subfigure}[b]{0.5\textwidth}
        \centering
        \includegraphics[trim={0 0 3.5cm 0}, clip, width=0.9\columnwidth]{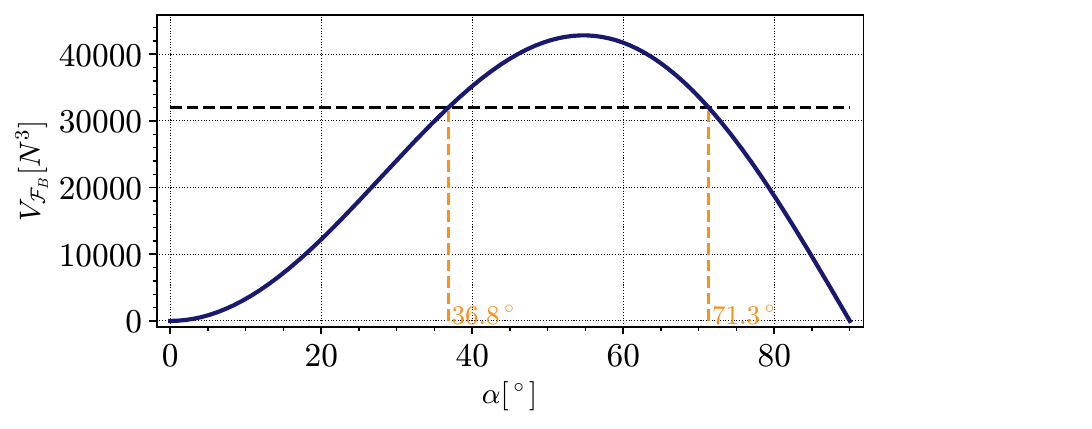}
        \caption{$V_{\mathcal{F}_B}$}
        \label{fig:V_FB_with_points}
    \end{subfigure}%
    \hfill%
    \begin{subfigure}[b]{0.3\textwidth}
        \centering
        \includegraphics[width=4cm]{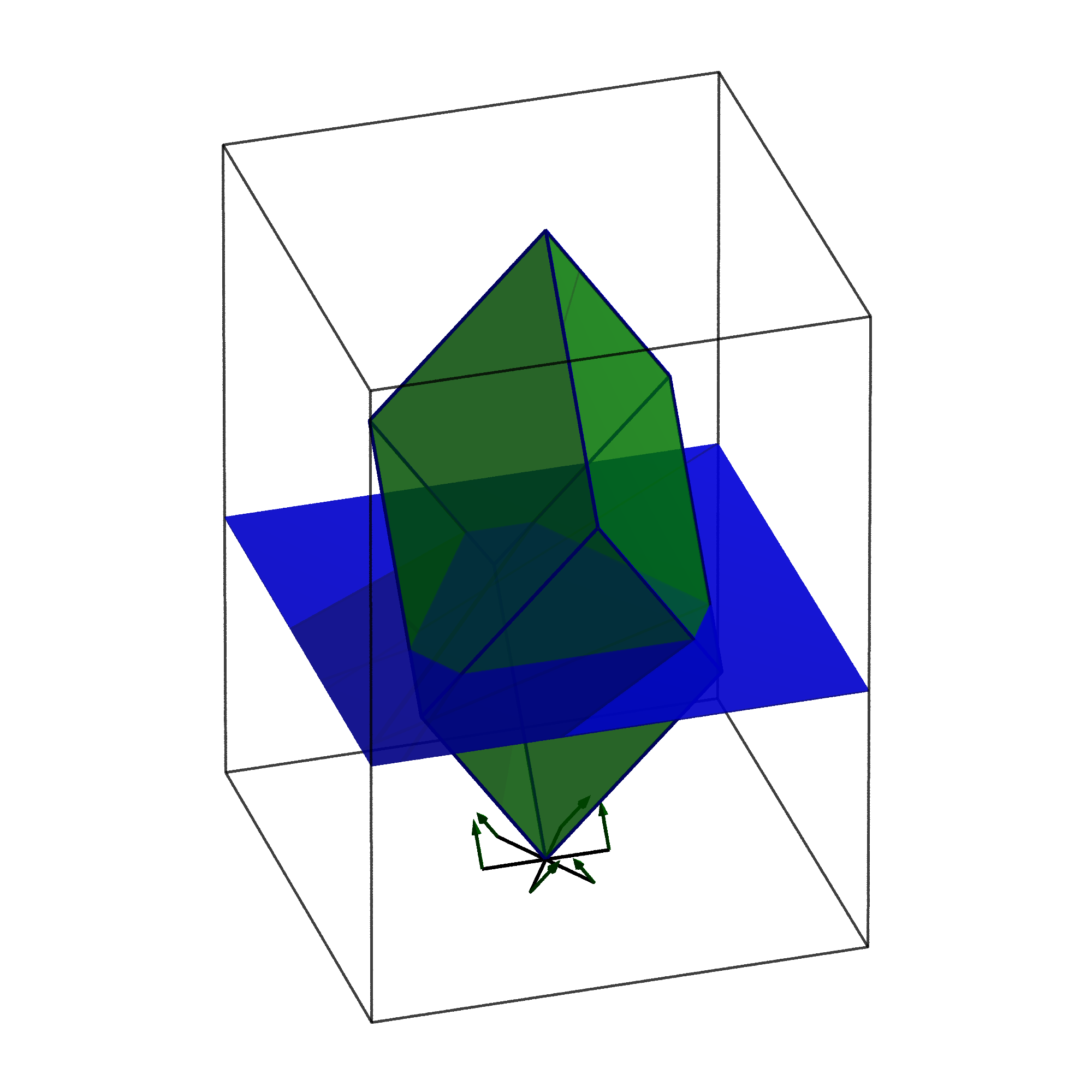}
        \caption{\centering $\alpha=36.8^\circ$ $V_{\mathcal{F}_B}=32000[N^3]$}
        \label{fig:poly_45}
    \end{subfigure}%
    \hfill%
    \begin{subfigure}[b]{0.2\textwidth}
        \centering
        \includegraphics[width=3.1cm]{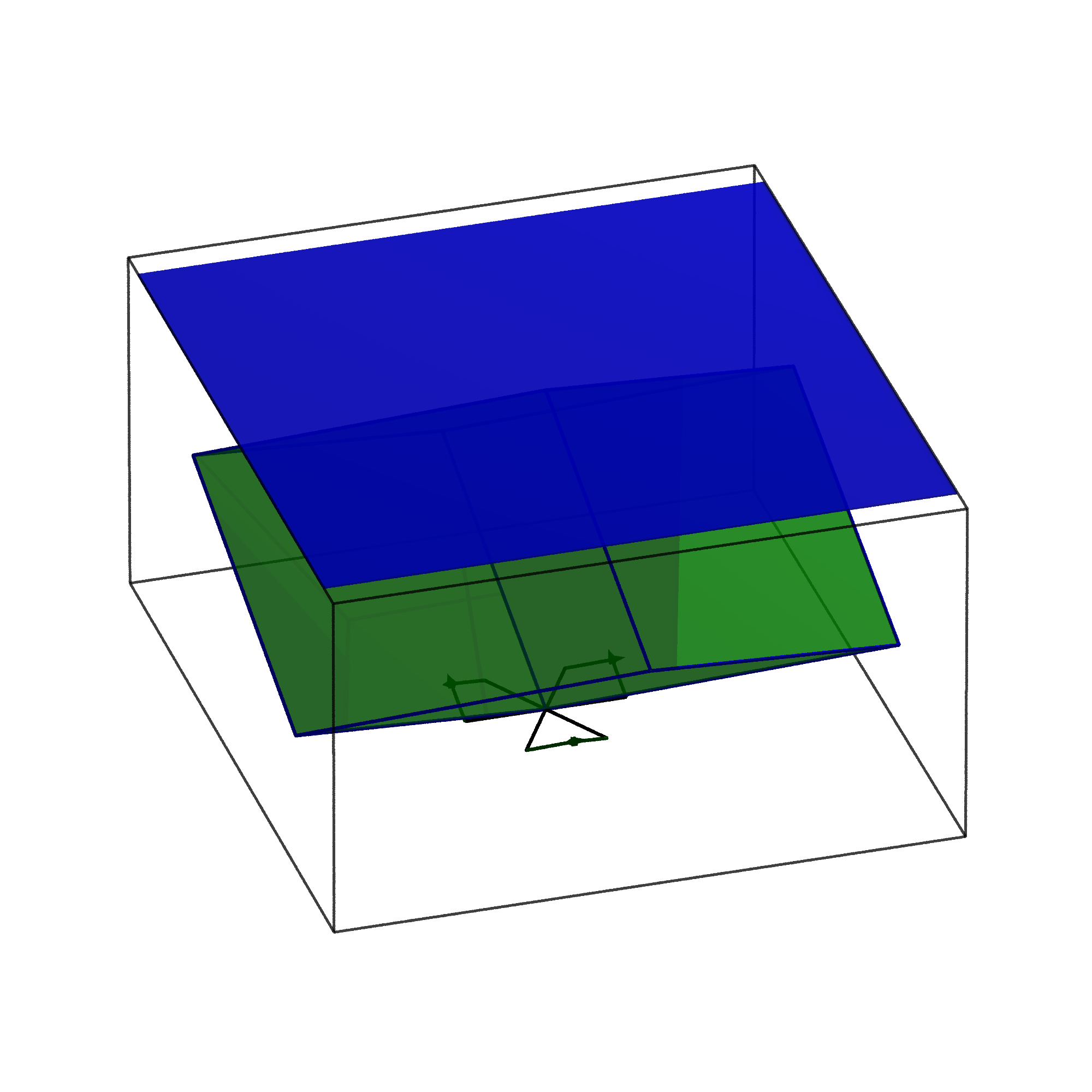}
        \caption{\centering $\alpha=71.3^\circ$ $V_{\mathcal{F}_B}=32000 [N^3]$}
        \label{fig:poly_66}
    \end{subfigure}%
    \hfill
    \caption{Investigation of $V_{\mathcal{F}_B}$: (a) value of $V_{\mathcal{F}_B}$ in function of $\alpha \in \Gamma_\alpha$, 
    (b-c) 3D polytopes representing the zero-moment force spaces characterized by the same volume (related to the tilt angles marked in orange in Fig.~\ref{fig:V_FB_with_points})  - 
     the blue plane represents the hovering plane introduced in Sec.~\ref{sec:gravit_compensation_constraint}.
    }
\end{figure*}

\subsection{Zero-moment Control Force Volume}

To investigate the zero-moment control force space, we first observe that any  $\mathbf{f}_c^B \in \mathcal{F}_B$ can be expressed as $\mathbf{f}_c^B = \mathbf{F}_\alpha \mathbf{B}_\alpha \tilde{\mathbf{u}}_B$ with $\tilde{\mathbf{u}}_B = \colvec{\tilde{u}_{B,1} \;
\tilde{u}_{B,2} \;
\tilde{u}_{B,3}}^\top \in \bar{\mathcal{U}}^3$. In addition, exploiting~\eqref{eq:FM}, we verify that $\mathbf{B}_\alpha = \colvec{\mathbf{I}_3 & \mathbf{I}_3}^\top$ for any $\alpha \in \Gamma_\alpha$. This implies  that
for any $\mathbf{u}_B \in \mathcal{U}_B$ it is $u_{B,k} = u_{B,k+3} = \tilde{u}_{B,k}$ with $\tilde{u}_{B,k} \in \bar{\mathcal U}$ being $k \in \{1,2,3\}$. More interestingly, by introducing the matrix $\mathbf{H}_\alpha=\mathbf{F}_\alpha\mathbf{B}_\alpha  \in \mathbb{R}^{3 \times 3}$, we have that any  $\mathbf{f}_c^B \in \mathcal{F}_B$ is such that
\begin{align}
\label{eq:alpha_force}
 \mathbf{f}_c^B  &=\mathbf{H}_\alpha \tilde{\mathbf{u}}_B = c_f \colvec{0 & \sqrt{3} s\alpha & -\sqrt{3} s\alpha \\
2 s\alpha & - s\alpha & - s\alpha \\
2  c\alpha & 2  c\alpha & 2  c\alpha } \colvec{\tilde{u}_{B,1} \\
\tilde{u}_{B,2} \\
\tilde{u}_{B,3}}.
\end{align}

The zero-moment control force space is defined by the combination of the columns of the matrix $\mathbf{H}_\alpha$ multiplied by the maximum value assignable to any propeller control input, corresponding in this case to the maximum value assignable to any $\tilde{u}_{B,k}$ with  $k \in \{1,2,3\}$. Formally, it is  
\begin{align}
    \mathcal{F}_B 
    &= \left\{ \textstyle\sum_{i=1}^{3}  \epsilon_i \; \bar{u}{\mathbf{H}}_\alpha\mathbf{e}_i, \; \epsilon_i \in [0,1]
    \right\} \label{eq:Fspace_def_2}.
\end{align}
From a geometrical point of view, the expression~\eqref{eq:Fspace_def_2} of $\mathcal{F}_B$ corresponds to a bounded, convex, and finite polytope whose volume $V_{\mathcal{F}_B} \in \mathbb{R}_{\geq 0}$ constitutes a suitable index for the STHs maneuverability. This can be computed as
\begin{equation}
\label{eq:volumeF}
V_{\mathcal{F}_B}= \left|\text{det}\left(\bar{u}\mathbf{H}_\alpha \right)\right|.
\end{equation}
The determinant of the matrix $\bar{u}\mathbf{H}_\alpha$ equals the magnitude of the scalar triple product of its column vectors, which corresponds to the volume of the parallelepiped they span in 3D space.
The absolute value in~\eqref{eq:volumeF} ensures this volume is treated as positive.  Thus the index $V_{\mathcal{F}_B}$  represents the volume of the convex hull spanned by the columns of $\bar{u} \mathbf{H}_\alpha$. 
The index~\eqref{eq:volumeF} can be computed in closed form and depends on the actuator parameters and tilt angle selection as follows
\begin{equation}
\label{eq:volumeF_exp}
    V_{\mathcal{F}_B} = 12 \sqrt{3} (c_f \bar{u})^3  c\alpha  s\alpha^2.
\end{equation}

Fig.~\ref{fig:V_FB_with_points}  illustrates how the index~\eqref{eq:volumeF_exp} varies with the selection of $\alpha \in \Gamma_\alpha$.
The volume of the polytope corresponding to $\mathcal{F}_B$ 
decreases as the tilt angle approaches either $0^{\degree}$ or $90^{\degree}$, and increases as $\alpha$ assumes central values in $\Gamma_\alpha$. The curve shows asymmetry, peaking around $\alpha\sim 55^{\degree}$.
However, different choices of $\alpha$ can yield the same value of $V_{\mathcal{F}_B}$. Therefore, relying solely on this index provides an incomplete assessment of STH maneuverability. For instance, the 3D polytopes depicted in Fig.s~\ref{fig:poly_45}-\ref{fig:poly_66} represent the zero-moment force space for various $\alpha$ selections: although their volumes are roughly equivalent, their dimensional characteristics (height, width, depth) differ significantly. Thus, employing supplementary metrics is crucial for obtaining a deeper understanding of the impact of the $\alpha$ selection.


Aiming at investigating STH maneuverability for interactive tasks, we analyze static hovering conditions where UAVs must counteract gravity and reject moment disturbances. This necessity leads to a reduction in $\mathcal{F}_B$, and the magnitude of the resulting extra-hovering zero-moment force space is a valuable indicator of platform maneuverability. 

\subsection{Gravitation Compensation Constraint}
\label{sec:gravit_compensation_constraint}

First, we focus on the convex hull of the intersection between $\mathcal{F}_B$ and the plane perpendicular to $\mathbf{e}_3$ describing the control force space $\{ \mathbf{f}_c = \colvec{ f_c^x \; f_c^y \; f_c^z}^\top \in \mathbb{R}^3 \; \vert \; f_c^z = mg\}$ (hovering plane). More precisely, based on~\eqref{eq:alpha_force}, we investigate  $\text{conv}(\mathcal{F}_B^h)$, i.e., the convex hull of the set

\begin{align}
\label{eq:hovering_forces}
       \mathcal{F}_B^h &=\left\{ \mathbf{f}_c = c_f \colvec{\sqrt{3} s\alpha (\tilde{u}_{B,2}-\tilde{u}_{B,3}) \\
 s\alpha (2\tilde{u}_{B,1} - \tilde{u}_{B,2} - \tilde{u}_{B,3})\\
2  c\alpha (\tilde{u}_{B,1}+
\tilde{u}_{B,2}+
\tilde{u}_{B,3})},
             \right.\\ 
    & \hspace*{1cm}  \left. 
        \textstyle\sum_{k=1}^3 \tilde{u}_{B,k} = \frac{mg}{2 c_f c\alpha},\;  \tilde{u}_{B,k} \in \bar{\mathcal{U}} , k \in\{1,2,3\}\right\}. \nonumber
\end{align}

    \begin{table*}[t!]
    \centering
    \resizebox{0.99\textwidth}{!}{
        \begin{tabular}{|c|lll|llll|}
        \hline
\multirow{3}{*}{A}& $\tilde{u}_{B,1} =0 $ & $ \tilde{u}_{B,3} = 0$ & $\tilde{u}_{B,2} = \frac{mg}{2 c_f c\alpha}$
&  $f_{c,1}^x = \frac{\sqrt{3}}{2}mg \, t\alpha$ & ($f_{c,max}^x$) &$f_{c,1}^y = -\frac{mg}{2} \, t\alpha$ & ($f_{c,min}^y$)\\
& $\tilde{u}_{B,1} =0 $ & $  \tilde{u}_{B,2} = 0$ & $\tilde{u}_{B,3} = \frac{mg}{2 c_f c\alpha}$ & $f_{c,2}^x = -\frac{\sqrt{3}}{2}mg \, t\alpha$ & ($f_{c,min}^x$) & $f_{c,2}^y = -\frac{mg}{2} \, t\alpha$  & ($f_{c,min}^y$)\\
& $\tilde{u}_{B,2} =0 $ & $  \tilde{u}_{B,3} = 0$ & $\tilde{u}_{B,1} = \frac{mg}{2 c_f c\alpha}$
&  $f_{c,3}^x = 0$ & & $f_{c,3}^y = mg \, t\alpha$ & ($f_{c,max}^y$) \\ \hline
\multirow{6}{*}{B} &
 $\tilde{u}_{B,2} = \bar{u}$ & $\tilde{u}_{B,3} = 0$ & $\tilde{u}_{B,1} = \frac{mg}{2 c_f c\alpha}-\bar{u}$ & $f_{c,1}^x = \sqrt{3}c_f \bar{u} \,s\alpha $  &  ($f_{c,max}^x$) & $f_{c,1}^y  = mg \, t\alpha - 3c_f\bar{u}\, s\alpha $ & \\
&  $\tilde{u}_{B,2} = \bar{u}$ & $\tilde{u}_{B,1} = 0$ & $\tilde{u}_{B,3} = \frac{mg}{2 c_f c\alpha}-\bar{u}$  &  $f_{c,2}^x = \sqrt{3} \left(2c_f \bar{u} \, s\alpha  - \frac{mg}{2} t\alpha \right)$ & &$f_{c,2}^y = -\frac{mg}{2} \, t\alpha$ & ($f_{c,min}^y$)\\
&  $\tilde{u}_{B,3} = \bar{u}$ & $\tilde{u}_{B,2} = 0$ & $\tilde{u}_{B,1} = \frac{mg}{2 c_f c\alpha}-\bar{u}$  & $f_{c,3}^x = -\sqrt{3}c_f \bar{u} \, s\alpha $ & ($f_{c,min}^x$) & $f_{c,3}^y  = mg \, t\alpha - 3c_f \bar{u} \, s\alpha $  & \\
&  $\tilde{u}_{B,3} = \bar{u}$ & $\tilde{u}_{B,1} = 0$ & $\tilde{u}_{B,2} = \frac{mg}{2 c_f c\alpha}-\bar{u}$  & $f_{c,4}^x = -\sqrt{3} \left(2c_f \bar{u} \, s\alpha  - \frac{mg}{2} t\alpha \right)$ & & $f_{c,4}^y = -\frac{mg}{2} \, t\alpha$  &  ($f_{c,min}^y$)\\
 &  $\tilde{u}_{B,1} = \bar{u}$ & $\tilde{u}_{B,3} = 0$ & $\tilde{u}_{B,2} = \frac{mg}{2 c_f c\alpha}-\bar{u}$  &  $f_{c,5}^x = \sqrt{3} \left(\frac{mg}{2} t\alpha - c_f \bar{u} \, s\alpha  \right)$ & & $f_{c,5}^y = 3 c_f \bar{u} \, s\alpha  - \frac{mg}{2} t\alpha$  &  ($f_{c,max}^y$)\\
&  $\tilde{u}_{B,1} = \bar{u}$ & $\tilde{u}_{B,2} = 0$ & $\tilde{u}_{B,3} = \frac{mg}{2 c_f c\alpha}-\bar{u}$  & $f_{c,6}^x = -\sqrt{3} \left(\frac{mg}{2} t\alpha - c_f \bar{u} \, s\alpha  \right)$ & & $f_{c,6}^y =3 c_f \bar{u} \, s\alpha  - \frac{mg}{2} t\alpha$ & ($f_{c,max}^y$) \\ \hline
\multirow{3}{*}{C} & $\tilde{u}_{B,1} = \bar{u}$ & $\tilde{u}_{B,2} = \bar{u}$ & $\tilde{u}_{B,3} = \frac{mg}{2 c_f c\alpha}-2\bar{u}$ & $f_{c,1}^x = \sqrt{3} \left( 3c_f \bar{u} \, s\alpha  - \frac{mg}{2} t\alpha \right)$ &  ($f_{c,max}^x$) & $f_{c,1}^y = 3c_f \bar{u} \, s\alpha  -\frac{mg}{2} \, t\alpha$ & ($f_{c,max}^y$)\\
& $\tilde{u}_{B,2}  = \bar{u}$ & $ \tilde{u}_{B,3} = \bar{u}$ & $\tilde{u}_{B,1} = \frac{mg}{2 c_f c\alpha}-2\bar{u}$
& $f_{c,2}^x = 0$ &  & $f_{c,2}^y = mg \, t\alpha - 6c_f \bar{u} \, s\alpha $ &  ($f_{c,min}^y$)\\
& $\tilde{u}_{B,1}  = \bar{u}$ & $ \tilde{u}_{B,3} = \bar{u}$ & $\tilde{u}_{B,2} = \frac{mg}{2 c_f c\alpha}-2\bar{u}$ & $f_{c,3}^x = -\sqrt{3} \left( 3c_f \bar{u} \, s\alpha  - \frac{mg}{2} t\alpha \right)$ &  ($f_{c,min}^x$) & $f_{c,3}^y = 3c_f \bar{u} \, s\alpha  -\frac{mg}{2} \, t\alpha$ & ($f_{c,max}^y$)\\ \hline
        \end{tabular}
        }
        \caption{Vertices of the convex hull of $\mathcal{F}_B^h$  in  the cases A-C and the corresponding values of the coefficients $\tilde{u}_{B,k}$, $k=1,2,3$.}
        \label{tab:extreme}
    \end{table*}

Given these premises, to figure out the vertexes of $\text{conv}(\mathcal{F}_B^h)$, we investigate the actuators' extreme working conditions and we distinguish the  following cases:
\begin{itemize}
\item[A.] $\bar{u} \geq \frac{mg}{2 c_f c\alpha}$ then it is required the action of at least a pair of propellers to counteract the gravity;
\item[B.] $\frac{mg}{4 c_f c\alpha} \leq \bar{u} < \frac{mg}{2 c_f c\alpha}$  then it is required the action of at least two pairs of propellers to counteract the gravity;
\item[C.] $\frac{mg}{6 c_f c\alpha} \leq \bar{u} < \frac{mg}{4 c_f c\alpha}$  then it is required the action of all the propellers to counteract the gravity; 
\item[D.] $\bar{u} < \frac{mg}{6 c_f c\alpha}$ then the STH is not capable of counteract the gravity force and then to take off.
\end{itemize}
{Note that, based on the actuators parameters $c_f$ and $\bar{u}$, the selection of the tilt angle is constrained to different sub-intervals of $\Gamma_\alpha$ corresponding to cases A-D.}

Focusing on the cases A-C, we verify that the actuators extreme working conditions correspond to the selection of the input parameters reported in Tab.~\ref{tab:extreme}. In its last column, we provide the corresponding expressions for the $f_c^x$ and $f_c^y$ control force components,
indicating also whether these values correspond to a minimum/maximum in the light of~\eqref{eq:hovering_forces}.
Thus, it is possible to evaluate the area ${A}_{\mathcal{F}_B^h} \in \mathbb{R}_{\geq{0}}$ of the resulting convex hull $\text{conv}(\mathcal{F}_B^h)$ for the cases A-C. In detail, exploiting the shoelace/Gauss's area formula, the value of ${A}_{\mathcal{F}_B^h}$ can be computed as
\begin{equation}
\label{eq:area}
{A}_{\mathcal{F}_B^h} = \frac {1}{2}\left|\sum _{j=1}^{n_\bullet}(f_{c,j}^xf_{c,j+1 (\text{mod} n_\bullet)}^y-f_{c,j+1 (\text{mod} n_\bullet)}^xf_{c,j}^y)\right|
 \end{equation}
 where $n_\bullet \in \mathbb{N}$, with $\bullet= \{ A, B, C\}$, denotes the number of vertexes in the considered cases ($n_A = n_C = 3$ and $n_B=6)$. It follows that 
\begin{equation}
\label{eq:area_exp}
   {A}_{\mathcal{F}_B^h} = 
    \begin{cases}
     \frac{3\sqrt{3}}{4} \left( mg \, t\alpha \right)^2 & \text{A.}\\
   \frac{3\sqrt{3}}{4} \left( mg \, t\alpha \right)^2 - 9\sqrt{3} \left( \frac{mg}{2} t\alpha - c_f \bar{u} \, s\alpha  \right)^2 & \text{B.}\\
 3\sqrt{3} \left(3c_f \bar{u} \, s\alpha  - \frac{mg}{2} t\alpha \right)^2 & \text{C.}\\
   \end{cases} 
\end{equation}

\begin{figure}[t!]
    \centering
\includegraphics[trim={0 0 3.5cm 0},clip,width=0.95\columnwidth]{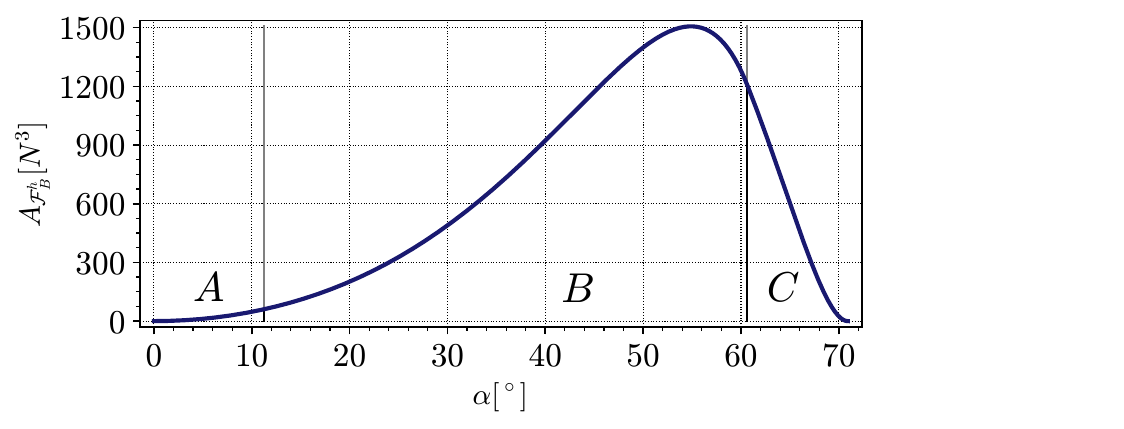}
    \caption{Value of ${A}_{\mathcal{F}_B^h}$ in function of $\alpha \in \Gamma_\alpha$, highlighting cases A-C.
    }
    \label{fig:area}
\end{figure}

Fig.~\ref{fig:area} illustrates how ${A}_{\mathcal{F}_B^h}$ varies based on the selection of $\alpha \in \Gamma_\alpha$. The asymmetry of the curve is more evident, as compared to Fig.~\ref{fig:V_FB_with_points}. Furthermore, even though the peak of the curve occurs at approximately the same angle as for the volume $V_{\mathcal{F}_B}$, the maximum angle that can be imposed is more stringent, about $20^\circ$.

\begin{figure*}[t!]
\centering
 \hfill
 \begin{subcaptionblock}{.29\textwidth}
     \centering
     \includegraphics[height=5.1cm]{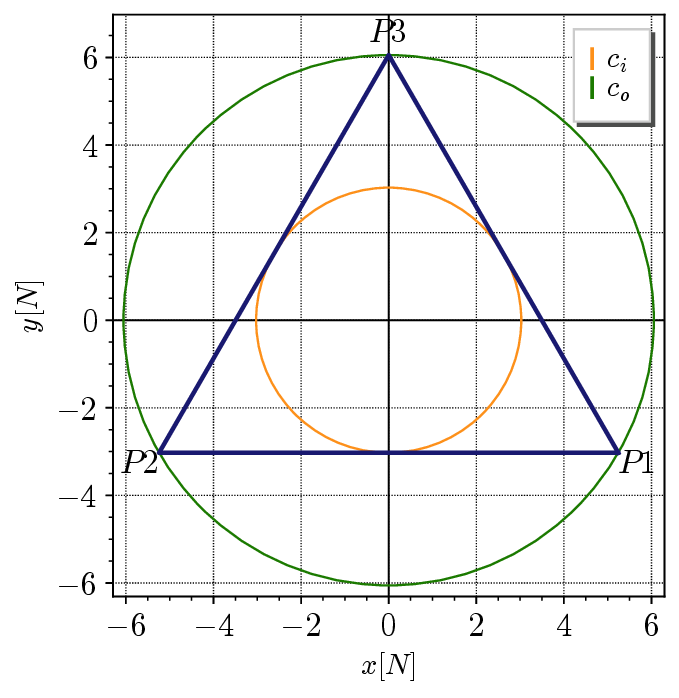}
     \caption{ $\alpha = 10^{\degree}$ (case A)}\label{fig:area_circles_case_A}
 \end{subcaptionblock}%
 \hfill
 \begin{subcaptionblock}{.33\textwidth}
     \centering
     \includegraphics[height=5cm]{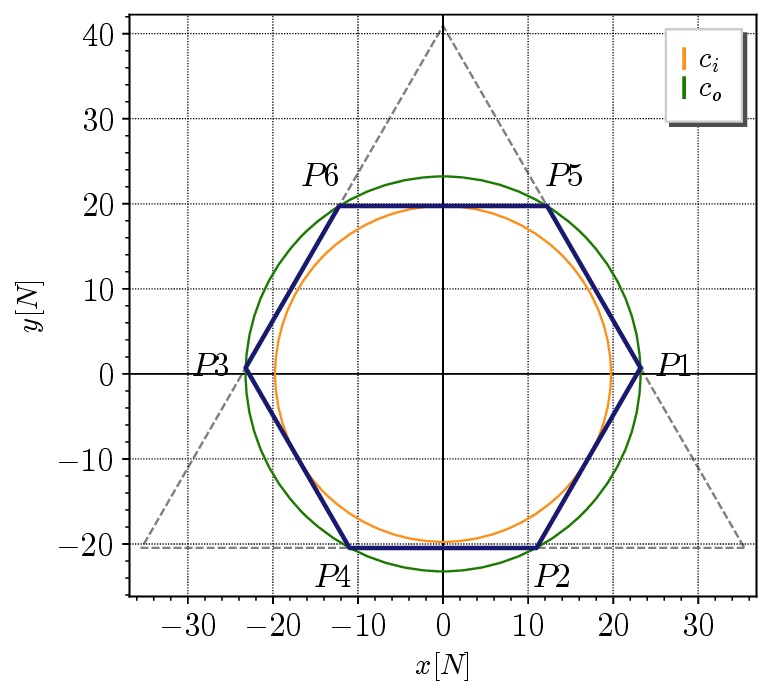}
     \caption{$\alpha = 50^{\degree}$ (case B)}\label{fig:area_circles_case_B}
 \end{subcaptionblock}%
 \hfill
 \begin{subcaptionblock}{.34\textwidth}
     \centering
     \includegraphics[height=5cm]{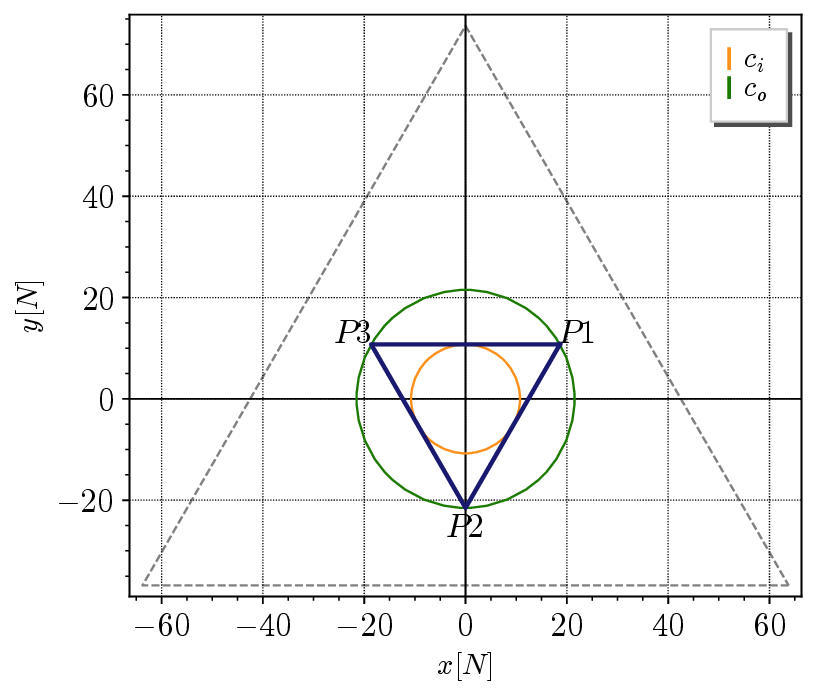}
     \caption{$\alpha = 65^{\degree}$ (case C)}\label{fig:area_circles_case_C}
 \end{subcaptionblock}
 \hfill
\caption{
For $\alpha$ selection in the cases A-C,  representation of the inner $c_i$ and outer $c_o$ circles and of the convex hull of $\mathcal{F}_B^h$ delimited by $P_i$, $i \in \{1,\ldots,6\}$ --- corresponding to the extreme force components in Tab.~\ref{tab:extreme} --- denoting the forces that guarantee the UAV can hover.}
\label{fig:area_circles}
\end{figure*}

To deeply understand the maneuverability of STHs, we also investigate the radii of the outer (circumscribed) and inner (inscribed) circles of the convex hull $\text{conv}(\mathcal{F}_B^h)$. The radius $r_{o} \in \mathbb{R}_{\geq 0}$ of the outer circle quantifies the aggressiveness of the platform, capturing the extremes of the achievable $f_x$ and $f_y$ force components. 
Conversely, the radius $r_{i} \in \mathbb{R}_{\geq 0}$ of the inner circle serves as an index of the platform robustness, reflecting the ability to generate equal force magnitudes along the $x$ and $y$ axes of $\mathscr{F}_B$. 
In all three cases mentioned, these circles are centered at the origin of the $(f_c^x,f_c^y)$-plane. Their radii can be determined using the coordinates of the extreme points listed in Tab.~\ref{tab:extreme}. Specifically, they result
\begin{itemize}
\item[A.] $r_{i} = |f_{c,1}^y|$ and $r_{o} = |f_{c,3}^y|$,
\item[B.] $r_{i} =\min \left\lbrace |f_{c,2}^y|, |f_{c,5}^y| \right\rbrace$ and $r_{o} = \vert \colvec{ f_{c,1}^x \;\; f_{c,1}^y }\vert$,
\item[C.] $r_{i} = |f_{c,1}^y|$ and $r_{o} = |f_{c,2}^y|$.
\end{itemize}

In Fig.~\ref{fig:area_circles}, for each of the three cases, the convex hull of $\mathcal{F}_B^h$ and the corresponding inner and outer circles are depicted. In cases A and C, $\text{conv}(\mathcal{F}_B^h)$ forms an equilateral triangle. Conversely, in case B, $\text{conv}(\mathcal{F}_B^h)$ is a hexagon with vertices lying on the perimeter of the triangle depicted in case A. As the parameter $\alpha$ increases, these vertices gradually move away from the vertices of the original triangle, eventually aligning with those of the triangle depicted in case C.

\begin{figure}[t!]
\centering
\includegraphics[trim={0 0 0cm 0}, clip, width=0.95\columnwidth]{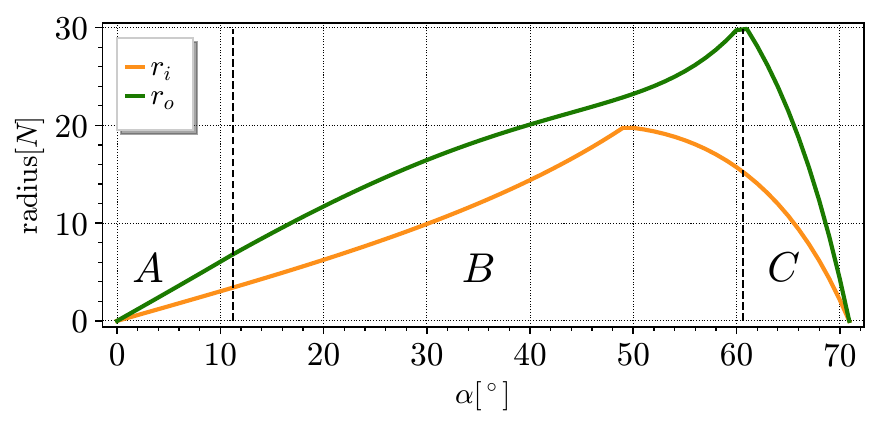}
\caption{Value of $r_i$ and $r_o$ in function of $\alpha \in \Gamma_\alpha$, highlighting cases A-C.}
\label{fig:radii}
\end{figure}

In Fig. \ref{fig:radii}, we illustrate how the radii of the inner and outer circles vary as functions of the angle $\alpha$. The outer circle's radius attains its maximum value when cases B and C coincide; at this point, $\text{conv}(\mathcal{F}_B^h)$ is a triangle with the vertices on the perimeter of the triangle introduced in the case A. Concerning the inner circle's radius, it reaches its maximum in case B, where $\text{conv}(\mathcal{F}_B^h)$ is a regular hexagon.

\subsection{Extra-hovering Control Force Volume}

Finally, we investigate the volume $V_{\mathcal{F}_B^h} \in \mathbb{R}_{\geq0}$ of the space of the forces exceeding the gravity compensation required by the hovering conditions. Formally, we focus on the space $\{ \mathbf{f}_c = \colvec{ f_c^x \; f_c^y \; f_c^z}^\top \in \mathcal{F}_B  \; \vert \; f_c^z \geq mg\}$. 

\begin{figure*}[t!]
\centering
 \hfill
 \begin{subcaptionblock}{.32\textwidth}
     \centering
     \includegraphics[trim={12cm 5cm 0cm 0}, clip, height=6cm]{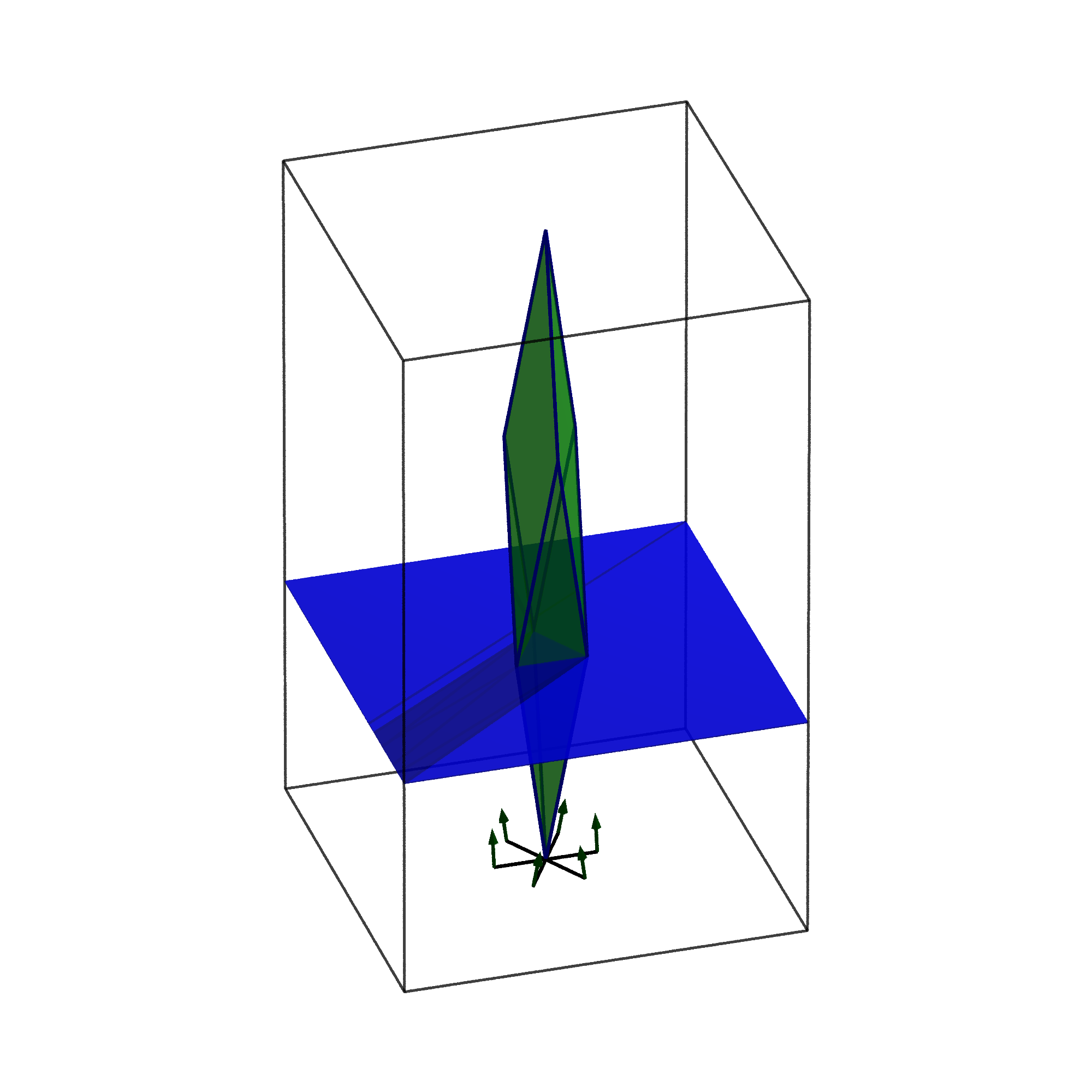}
     \caption{ $\alpha = 10^{\degree}$ (case A)}\label{fig:gems_plot_case_A}
 \end{subcaptionblock}%
 \hfill
 \begin{subcaptionblock}{.32\textwidth}
     \centering
     \includegraphics[trim={0cm 0cm 0cm 0},clip,height=4.7cm]{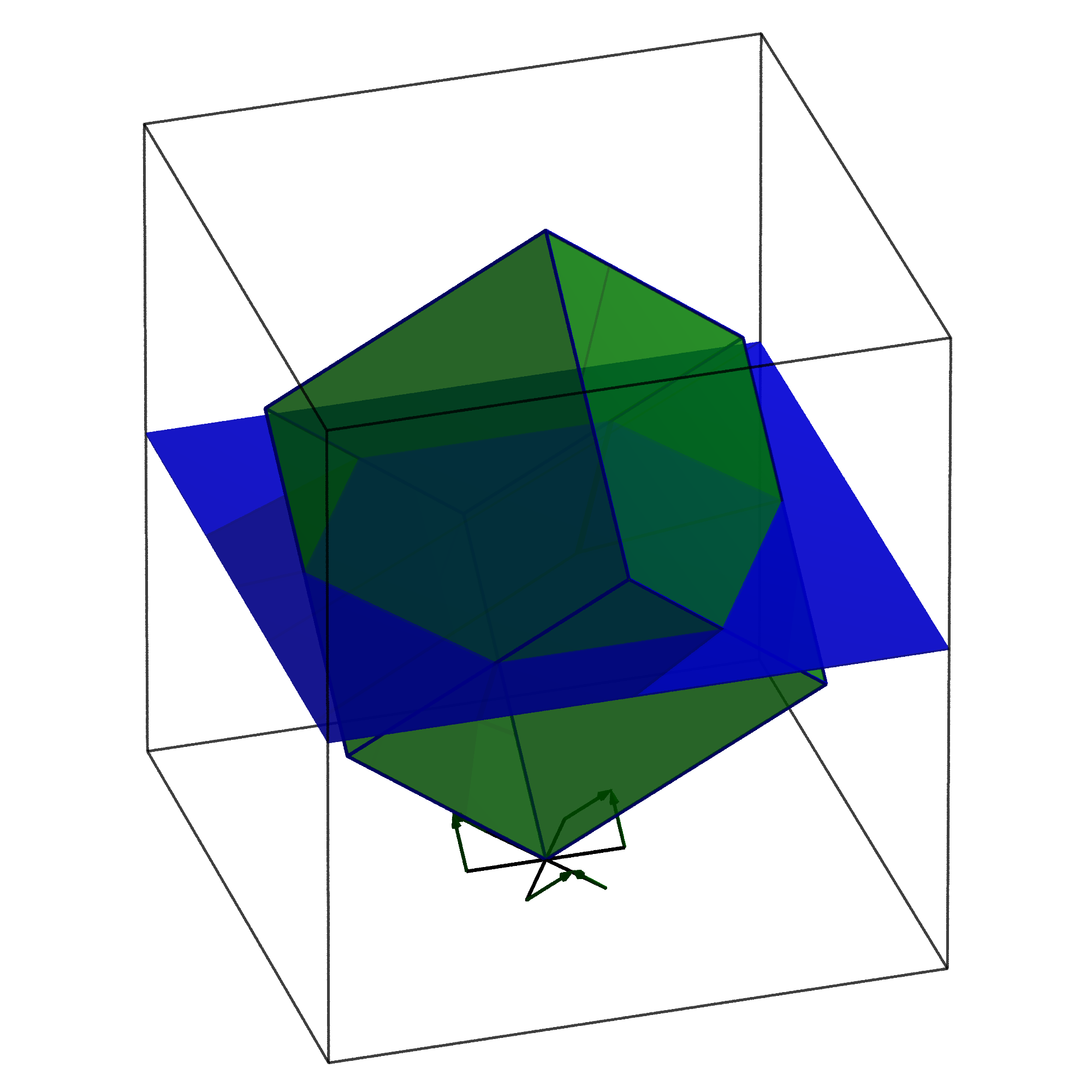}
     \caption{$\alpha = 50^{\degree}$ (case B)}\label{fig:gems_plot_case_B}
 \end{subcaptionblock}%
 \hfill
 \begin{subcaptionblock}{.32\textwidth}
     \centering
     \includegraphics[trim={0cm 4cm 0cm 0},clip,height=4.2cm]{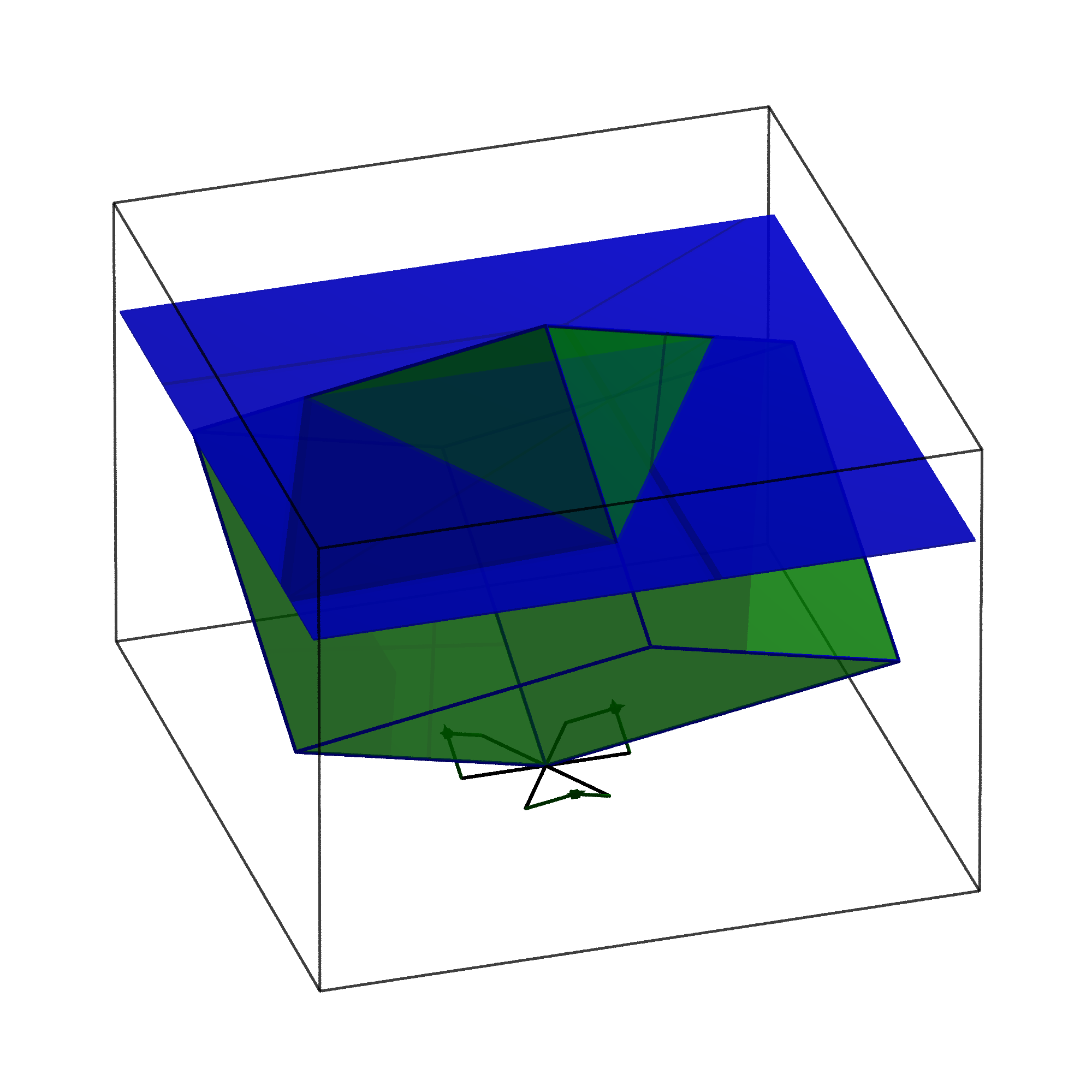}
     \caption{$\alpha = 65^{\degree}$ (case C)}\label{fig:gems_plot_case_C}
 \end{subcaptionblock}
 \hfill
\caption{Extra-hovering control force space for various $\alpha$ selection in the cases A,B,C. On the bottom, a representation of the frame of the UAV, with green arrows representing the forces for each propeller. The blue plane represents the hovering plane $\mathcal{F}_B^h$  }
\label{fig:extra_volume}
\end{figure*}

Similarly to the previous indexes, we have to distinguish between the cases A-C introduced in Sec. \ref{sec:gravit_compensation_constraint}. For various values of $\alpha$ corresponding to the distinct cases, examples of the extra-hovering control force space are illustrated in Fig.~\ref{fig:extra_volume}, where the polytope representing the zero-moment control force space (depicted in green) intersects the hovering plane (illustrated in blue).
In all three cases, the volume $V_{\mathcal{F}_B^h}$ can be deduced through geometric intuition.

In case A, the volume $V_{\mathcal{F}_B^h} $ can be computed by subtracting from the polytope representing the zero-moment control force space $V_{\mathcal{F}_B}$  the portion of the same polytope under the hovering plane. We observe that this turns out to be a pyramid having a height equal to $mg$ and $A_{\mathcal{F}_B}^h$ as the base. Therefore, in case A, it is $V_{\mathcal{F}_B^h} = V_{\mathcal{F}_B} - \frac{1}{3} A_{\mathcal{F}_B}^h mg$.

\begin{figure}[t!]
\centering
\vspace{-0.5cm}
\hfill%
\resizebox{\columnwidth}{!}{\input{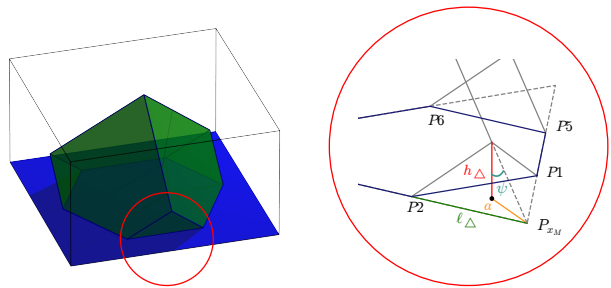}}
\hfill
\caption{Case B: polytope representing the extra-hovering control force space for $\alpha = 25^\circ$}
\label{fig:extra_hovering_polytope}
\end{figure}

In case B, it is observed that the polytope representing the extra-hovering control force space is geometrically characterized by a pyramid $\overline{\mathcal{P}}$, from which three smaller identical pyramids are subtracted (see left of Fig.~\ref{fig:extra_hovering_polytope}). Subsequently, as the rationale applies uniformly to any of the smaller pyramids, we denote one of them as $\underline{\mathcal{P}}$.
To compute the volume of such a polytope, we first compute the height of $\overline{\mathcal{P}}$ as $h_{+} = h_{max} - mg$, being $h_{max} = (\mathbf{e}_3^\top \mathbf{F}_\alpha)   (\bar u\mathds{1}_6) = 6 c_f c\alpha \bar u$ (with $\mathds{1}_n$ intended as one vector of dimension $n$) the maximum realizable force along $z$-axis by spinning all of the propellers at their maximum speed. Then, relying on the notation introduced in Fig.~\ref{fig:extra_hovering_polytope} on the right,  
we account for the equilateral triangle (due to the symmetries in the UAV geometry) formed by $P1, P2, P_{x_M}$  where $P_{x_M}$ is the extremum along the $x$-axis of the dashed triangle.
This is the basis of the pyramid $\underline{\mathcal{P}}$, whose height is  $h_\triangle = h_{+} - 2 c\alpha c_f \bar u$. Note that the second term of the subtraction corresponds to the projection along the $z$-axis of one of the edges of the entire polytope.
In detail, this edge measures $2 c_f \bar u$, corresponding to the norm of the vectors which generate $\mathcal{F}_B$.
To calculate the area of the basis of  $\underline{\mathcal{P}}$, it is suitable to introduce the angle $\psi \in [0,180^\circ]$ between the aforementioned edge and $h_\triangle$.
Henceforth, we easily compute the length of the segment  $a$, namely $\bar a = h_\triangle \tan\psi$, and the side of the equilateral triangle, namely $\ell_\triangle = 2 sin(60^\circ) a$. Thereafter, the volume of $\underline{\mathcal{P}}$ results to be $V_{\triangle} = \frac{1}{3} A_\triangle h_\triangle = \frac{1}{3} (\frac{\sqrt{3}}{4} \ell_{\triangle}^2) h_\triangle$.
On the other hand, the volume of  $\overline{\mathcal{P}}$ is equal to $V_{\overline{\mathcal{P}}} = \frac{1}{3} ({A}_{\mathcal{F}_B^h} + 3 A_\triangle) h_{+}$, given that its basis area results from the sum of ${A}_{\mathcal{F}_B}^h$ and the area of the bases of the three minor pyramids. In conclusion, in case B, it turns out that $V_{\mathcal{F}_B^h} = \frac{1}{3} ({A}_{\mathcal{F}_B^h} + 3 A_\triangle) h_{+}- 3 V_{\triangle}$.

In case C, the volume of the polytope representing the extra-hovering control force space can be derived from the previous case B by considering only the pyramid $\overline{\mathcal{P}}$. Therefore, in case C, it is $V_{\mathcal{F}_B^h} = \frac{1}{3} {A}_{\mathcal{F}_B^h} h_{+}$.

To sum up, the volume $V_{\mathcal{F}_B^h}$ is computed as follows:
\begin{equation}
\label{eq:extraVolume}
   {V}_{\mathcal{F}_B^h} = 
    \begin{cases}
      V_{\mathcal{F}_B} - \frac{1}{3} A_{\mathcal{F}_B}^h mg  & \text{A.}\\
   \frac{1}{3} {({A}}_{\mathcal{F}_B^h} + \frac{3 \sqrt{3}}{4} \ell_{\triangle}^2) h_{+} - (\frac{\sqrt{3}}{4} \ell_{\triangle}^2) h_\triangle & \text{B.}\\
    \frac{1}{3} {A}_{\mathcal{F}_B^h} h_{+} & \text{C.}\\
   \end{cases} 
\end{equation}

\begin{figure}[t!]
    \centering
    \vspace{0.9cm}
    \includegraphics[trim={0 0 3cm 0},clip,width=0.95\columnwidth]{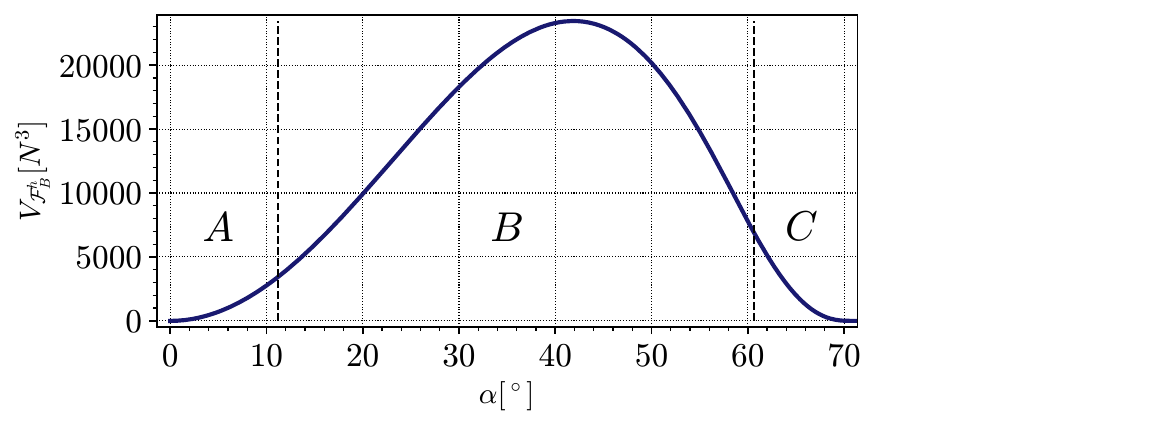}
    \caption{Value of ${V}_{\mathcal{F}_B^h}$ in function of $\alpha \in \Gamma_\alpha$, highlighting the cases A-C.}
    \label{fig:vol_hovering}
\end{figure}

Finally, we can observe the relationship between the overall hovering volume value and $\alpha$ in Fig. \ref{fig:vol_hovering}.
The peak of the curve occurs when $\alpha$ is approximately $42^{\degree}$ (case B). Consequently, $V_{\mathcal{F}_B^h}$ attains its maximum value at a lower $\alpha$ compared to the other metrics previously considered.

\section{Discussion}
\label{sec:discussion}
To enhance comprehension of the introduced metrics, we discuss their significance in relation to the angles that individually maximize them, as outlined in Tab.~\ref{tab:metrics}.  
This concluding section serves as an example of design analysis.

\begin{table}[h]
\centering
    \resizebox{0.9\columnwidth}{!}{
\begin{tabular}{cccccc}
\toprule
\multirow{2}{*}{Metrics} & \multicolumn{4}{c}{$\alpha \, [^\circ]$} \\
& 42 &  $49.5$ & $54.5$ & $55$ & $60.5$ \\
\midrule
$V_{\mathcal{F}_B} \, [N^3] $  
& $37039$ & $41802$ & $\mathbf{42843}$ & $42843$ & $41523$ \\
${A}_{\mathcal{F}_B^h} \, [N^2]$ 
& $1022$ & $1379$ & $1505$ & $\mathbf{1506}$ & $1222$ \\
$r_{o} \, [N]$  
& $20.71$ & $23.05$ & $25.26$ & $25.55$ & $\mathbf{30.34}$ \\
$r_{i} \, [N]$  
& $15.46$ & $\mathbf{19.81}$ & $18.66$ & $18.48$ & $15.34$ \\
$V_{\mathcal{F}_B^h} \, [N^3] $  
& $\mathbf{23450}$ & $20577$ & $15441$ & $14802$ & $7089$ \\
\bottomrule
\end{tabular}}

\caption{Proposed metrics for different values of $\alpha$ --- case B.}
\label{tab:metrics}
\end{table}

We observe that the index $V_{\mathcal{F}_B}$ reaches its maximum at $\alpha = 54.5^\circ$. This angle also nearly maximizes $A_{\mathcal{F}_B^h}$, suggesting the optimality of this angle selection for both metrics. 
However, we highlight that the proximity of these two maxima on the curve to the same angle is coincidental and specific to the parameters chosen in this case study.
Upon closer examination, there is a  notable disparity in the largest attainable radii. This inconsistency is evident when comparing with previous metrics and among the radii themselves.  Prioritizing robustness, a smaller $\alpha$ could allow for a higher force application along the hovering plane, although this comes at the cost of a reduced maximum achievable force with specific combinations of $f_x$ and $f_y$. 
This safer approach  is corroborated by the value attained by $V_{\mathcal{F}_B^h}$, which peaks at $42^\circ$, indicating enhanced hovering capabilities. These capabilities are valuable for both robustness and specific tasks that necessitate  force application in directions other than vertical. Therefore, for a conservative design strategy, the choice $\alpha= 50^\circ$ could represent a reasonable trade-off.


\section{Conclusions}
\label{sec:conclusions}

This paper presents a detailed analysis of the impact on the maneuverability of a STH of the tilt cant angles, namely the propellers tilt angles along the vehicle arms. Specifically, we focus on platforms whose rotors are alternately equally tilted.  For this class of UAVs, we investigate the dependence on the tilt angle $\alpha$ of the volume of the polytope of the feasible control forces ensuring torque decoupling up to gravity compensation. In doing so, we introduce geometric-inspired metrics that serve a dual purpose: influencing the resulting value of the volume under consideration and constituting indices of other STH properties such as aggressiveness and robustness. 
The outcome of this study is to offer valuable metrics for quantifying the maneuverability of existing STHs, while also providing guidelines for selecting the tilt angle during the platform design phase.

Future work will involve studying the impact of the dihedral angle, as well as exploring the combined role of both the cant and dihedral tilt angles. %

\bibliographystyle{IEEEtran}
\bibliography{bibliography}

\end{document}